\documentclass{WileyMSP-template}

\usepackage{xcolor}
\usepackage{graphicx}%
\usepackage{amsmath}%
\usepackage{ragged2e}
\justifying

\begin{document}




\title{2D silver-nanoplatelets metasurface for bright directional photoluminescence, designed with the local Kirchhoff's law}

\maketitle

\author{Elise Bailly}
\author{Jean-Paul Hugonin}
\author{Jean-René Coudevylle}
\author{Corentin Dabard}
\author{Sandrine Ithurria}
\author{Benjamin Vest}
\author{Jean-Jacques Greffet*}


\begin{affiliations}
E. Bailly, J.-P. Hugonin, B. Vest, J.-J. Greffet\\
Universit\'e Paris-Saclay, Institut d’Optique Graduate School, CNRS, Laboratoire Charles Fabry, 91120 Palaiseau, France\\
Email Address: jean-jacques.greffet@institutoptique.fr

J.-R. Coudevylle\\
Centre de Nanosciences et de Nanotechnologies, Universit\'e Paris-Saclay, CNRS, 91120 Palaiseau, France

C. Dabard, S. Ithurria\\
Laboratoire de Physique et d’Etude des Mat\'eriaux, ESPCI-Paris, PSL Research University, Sorbonne Universit\'e UPMC Univ Paris
06, CNRS, 10 Rue Vauquelin, 75005 Paris, France

\end{affiliations}



\begin{abstract}

Semiconductor colloidal nanocrystals are excellent light emitters in terms of efficiency and spectral control. Integrating them with a metasurface would pave the way to ultrathin photoluminescent devices with reduced amount of active material and performing complex functionalities such as beam shaping or polarization control. To design such a metasurface, a quantitative model of the emitted power is needed. Here, we report the design, fabrication and characterization of a $\approx 300$ nm thick light-emitting device combining a plasmonic metasurface with an ensemble of nanoplatelets. The source has been designed with a new methodology based on a local form of Kirchhoff's law. The source displays record high directionality and brightness. 



\end{abstract}


\section{Introduction}\label{sec1}
Light emitting diodes (LEDs) have revolutionized lighting and display thanks to their efficiency and compactness. However, controlling light properties such as directionality, spectrum or polarization requires additional optical elements, such as collimators, filters or polarizers. These components reduce the optical flux and increase the size of the system. Recent technological developments in nanophotonics have demonstrated the suitability of metasurfaces as tools to control light properties using a single multifunctional flat device. Initially, metasurfaces have been designed to shape a spatially and spectrally coherent wavefront \cite{Li18,Yu14,Yu11}. The building blocks of the metasurfaces behave as resonant antennas and the control on directionality is achieved via interferences of the scattered field thanks to the dephasing induced by the independent antennas. 

However, this approach is unsuitable for the creation of light-emitting metasurfaces composed of many incoherent emitters integrated in the metasurface. Indeed, the emitters produce a field spatially incoherent which cannot be diffracted in a well-defined direction. 

It is possible to overcome this issue by generating spatial coherence. Directional emission based on emitters deposited on a plasmonic or dielectric grating has been achieved with dye molecules \cite{Vecchi09,Steele10,Lozano13,Lozano14,WangS18}, fluorescent glass \cite{Vaskin18}, quantum wells \cite{DiMaria13,Iyer20,Heki22}, and more recently with semiconductor nanocrystals \cite{Rodriguez12,GuoTorma15,WangX20,Dabard23,Bossavit23}. 
Restoration of a spatially coherent field was done by coupling efficiently any emitter to a delocalized mode such as a surface plasmon, a surface phonon polariton or a guided wave. A metasurface consisting of  resonators can also be used if they interact so that surface lattice resonance modes exist \cite{GuoTorma15}. To emit light in a controlled way, periodic perturbations are introduced so that the surface modes are converted into leaky modes. 



This is the basic mechanism of directional sources of thermal radiation based on a grating ruled on a material supporting surface phonon polaritons as reported in reference \cite{Greffet02}. A review of the state of the art of light-emitting metasurfaces can be found in the following references \cite{Vaskin19,Lozano16}.

The formalism routinely used to simulate and compute light emission by an ensemble of fluorophores embedded or in the vicinity of a metasurface consists in modelling them as point dipoles, and summing incoherently the intensities emitted by each dipole separately \cite{Vaskin19}. This technique enables to account for the emission and polarization pattern due to the surface. However, it does not account in a self-consistent manner for the feedback of the presence of the layer of emitters on the metasurface resonances. In addition, the technique does not enable to optimize the emitted power as the dipole amplitude is unknown. In practice, the  power emitted by the fluorophores depends on the pumping intensity and also on their temperature. Recently, it has been shown that the emitted power by a semiconductor can be optimized by using a reciprocity argument \cite{Liu18, Heki22}. Two issues are still pending. There is no available upper bound of the emitted power so that there is no figure of merit to characterize how far from optimum is the design in terms of brightness. The second issue is that the available models do not capture the dependence of light emission properties on temperature nor on the pumping process of the emitters.

In contrast with fluorescent sources, thermal sources are designed using Kirchhoff's law which enables to optimize simultaneously not only directionality and polarization but also emitted power by designing emissivities approaching 1 \cite{Costantini15,Wojszvzyk21,Nguyen22,Nguyen23,Watts12,Baranov19,Overvig21,Arnold12,WangX23}.
Recent works have shown that the local Kirchoff's law can be successfully used to analyze complex photoluminescent emission situations \cite{Monin23,Bailly22,Perez23,Caillas21}. In this article, we show that the local form of Kirchhoff's law can be used to optimize the design of a light-emitting metasurface \cite{Greffet18}. We extend to photoluminescence what was first achieved in the context of thermal emission, with an innovative method of designing directional photoluminescent metasurface via the local Kirchoff's law. We then demonstrate a bright and directional photoluminescent source. 



\section{Experimental platform}\label{sec2}
\begin{figure}[h!]
\centering
\includegraphics[width = 0.5\textwidth]{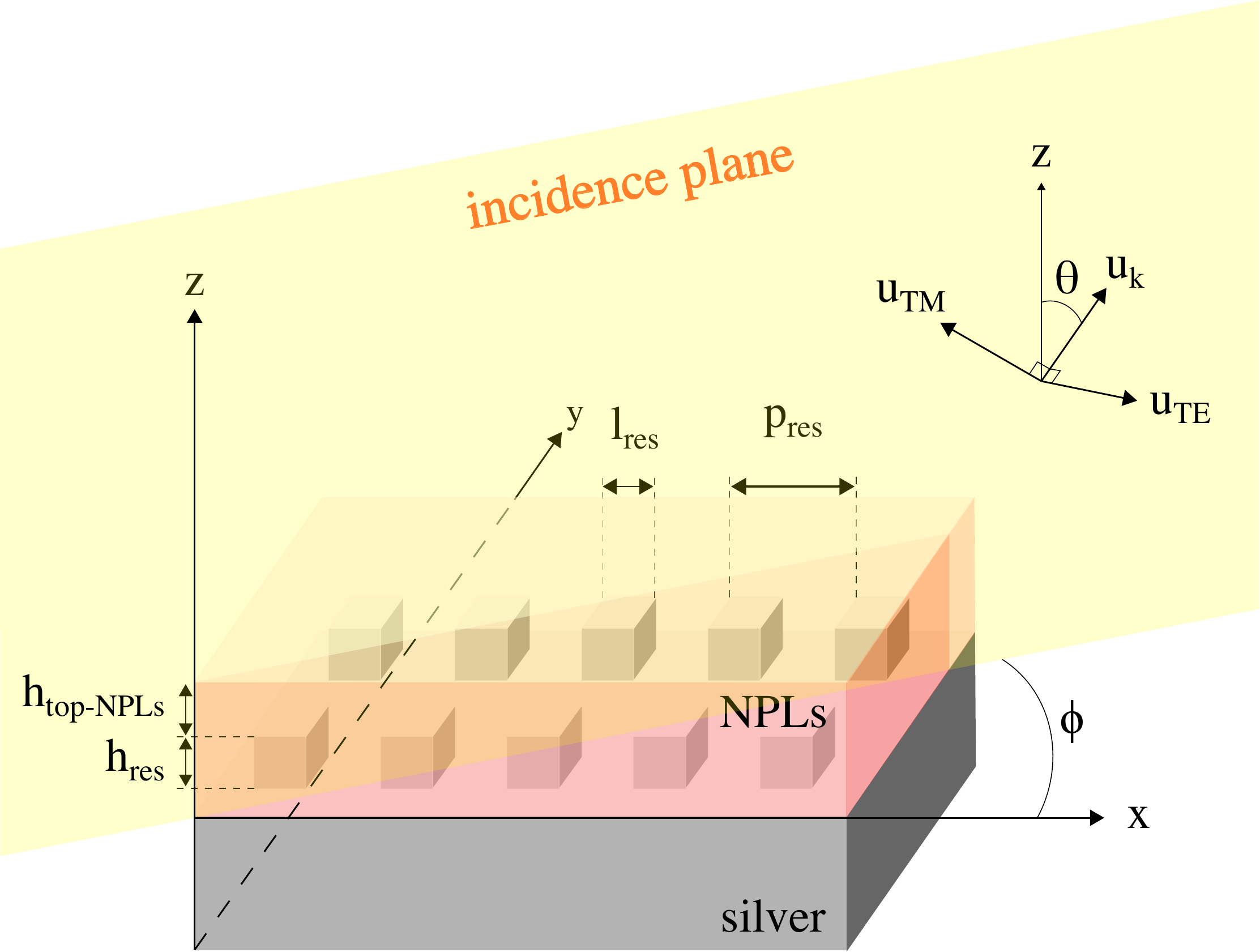}
\caption{Schematic view of the 2D photoluminescent metasurface composed of nanoplatelets (NPLs) on top of a 2D silver grating, with $h_{\mathrm{res}} = 100$ nm, $l_{\mathrm{res}} =$ 450 nm, $p_{\mathrm{res}} = $ 600 nm, $h_{\mathrm{top-NPLs}} =$ 2 nm. The geometric parameters are chosen in order to have a directional absorptivity (therefore reciprocally directional emissivity). The vectors $\mathbf{u_{\mathrm{TM}}}$ and $\mathbf{u_{\mathrm{TE}}}$ are unitary vectors for the Tranverse Magnetic (TM) and Transverse Electric (TE) directions.}
\label{schema}
\end{figure}

A schematic diagram of the light-emitting metasurface is shown in Figure \ref{schema}. We envision a metallic metasurface enabling the existence of surface plasmon polaritons which induce spatial coherence in the region of the interface \cite{Bailly21}. The directionality can then be obtained by coupling the surface mode to the far field by means of a grating \cite{Greffet02}. The whole surface is covered with emitters filling the grooves of the grating. Silver is used both as the substrate and grating material, because of its low losses in the visible spectrum. We use nanoplatelets (NPLs), because they are known to be bright emitters. NPLs are chemically synthesized core/shell semiconductor colloidal quantum wells. They consist of a 4 monolayers thick CdSe core (a monolayer is a stack of a cationic plane and an anionic plane \cite{Diroll23}) and a 4 monolayers thick ZnS shell, with oleic acid ligands attached to the surface. This leads to a thickness of approximatively 4 nm and an average lateral extension of 13 nm × 25 nm for each NPL. Their quantum yield has been measured and is equal to 50-55 $\%$. In the literature, a quantum yield close to unity has been obtained with core/shell CdSe/CdZnS NPLs using hot injection procedure \cite{Altintas19,Rossinelli17}. NPLs have two benefits: they display a reduced inhomogeneous broadening and they are robust to photobleaching and blinking compared to dye molecules \cite{Diroll23}. All these properties, coupled to their low production cost, make it an interesting choice of emitter for bright light emitting devices \cite{Baruj23}. The NPLs peak emission lies at 605 nm. Their absorption and emission spectra are given in the Supplementary Material \cite{Supplementary}.

\section{Numerical design of directional photoluminescent metasurfaces with the local Kirchhoff's law}\label{sec2}

In order to optimize the design a metasurface emitting light within a narrow solid angle around the direction normal to its surface, a model of its emission is needed. Here, we use the local Kirchhoff's law \cite{Greffet18,Rytov89,Wurfel82}. It is based on fluctuational electrodynamics and it has been shown that it enables to model electroluminescence as well as photoluminescence \cite{WangH18,Monin23,Perez23} by arbitrary resonant structures containing active regions. The pumping intensity of the active regions is characterized by a photon chemical potential $\mu$. In the specific case of electroluminescence, $\mu$ is given by the difference of the quasi-Fermi levels in the conduction and valence bands equal to $eV$, with $e$ the electron charge and $V$ the voltage bias, connected to the electrical intensity through the device by the characteristic $I(V)$ curve. 

The local Kirchhoff's law \cite{Greffet18} states that the power $\mathrm{d}^{2}P_{e}$ emitted by a layer of thermalized emitters at a wavelength $\lambda$ in the direction $\mathbf{u}$ and for a given polarization $l$ can be cast in the form:

\begin{equation}
    \mathrm{d}^{2}P_{e}^{(l)}(\mathbf{u},\lambda,\mu) = d\lambda  d\Omega\int_{V}d^{3}\mathbf{r}'\alpha^{(l)}(-\mathbf{u},\mathbf{r}',\lambda,\mu)\frac{hc^{2}}{\lambda^{5}}\frac{1}{\exp(\frac{hc}{\lambda k_{B}T}-\frac{\mu}{k_{B}T})-1},
    \label{eq_Kirchhoff}
\end{equation}

where $\alpha^{(l)}(-\mathbf{u},\mathbf{r'},\lambda)\varphi_{\mathrm{inc}}\mathrm{d}^{3}\mathbf{r'}$ is the absorption in a volume element $\mathrm{d}^{3}\mathbf{r'}$ of the \textit{active material} illuminated by an incident plane wave in the $-\mathbf{u}$ direction with a given polarization $l$ and a power flux per unit area $\varphi_{\mathrm{inc}}$. $T$ is the temperature and $\mu$ is the photon chemical potential. Note that Equation \eqref{eq_Kirchhoff} displays explicitly a temperature dependence, as well as a pumping intensity dependence via the chemical potential $\mu$. \\

For a flat source with area $S$, the emitted power can be expressed using the radiance $\mathcal{L}(\mathbf{u},\lambda)$:

\begin{equation}
   d^{2}P_{e}^{(l)}(\mathbf{u},\lambda) = d\lambda  d\Omega S \cos{\theta}\mathcal{L}(\mathbf{u},\lambda).
    \label{eq_radiometry}
\end{equation}

Assuming homogeneous temperature and chemical potential in the medium, these two equations lead to 
\begin{equation}
   \mathcal{L}^{(l)}(\mathbf{u},\lambda,\mu,T)  =  \mathcal{A}^{(l)}(-\mathbf{u},\lambda,\mu)\mathcal{L}_{\mathrm{BB}}(\lambda,\mu,T), 
    \label{C1_eq_abs}
\end{equation}
where $\mathcal{A}^{(l)}(-\mathbf{u},\lambda,\mu)$ is the absorptivity (which has an upper bound of 1), and $\mathcal{L}_{\mathrm{BB}}(\lambda,\mu,T) = \frac{hc^{2}}{\lambda^{5}}\frac{1}{\exp(\frac{hc}{\lambda k_{B}T}-\frac{\mu}{k_{B}T})-1}$ is the generalized Planck's law. In an aside, we would like to point out the fact that $\mathcal{L}_{\mathrm{BB}}(\lambda,\mu = 0,T)$ corresponds to Planck's law i.e. the blackbody radiance. Moreover, we emphasize that the absorptivity $\mathcal{A}(-\mathbf{u},\lambda,\mu)$ must not be computed in the entire structure but only in the active region. Thus, it cannot be retrieved from the reflection and transmission factors. \\
Since the light directionality is given by the angular dependence of the radiance $\mathcal{L}^{(l)}(\mathbf{u},\lambda,\mu,T)$, Equation \eqref{C1_eq_abs} explicitly shows that it matches the angular width of the absorptivity function $\mathcal{A}^{(l)}(-\mathbf{u},\lambda,\mu,T)$, showing that the absorptivity is the right figure of merit for light emission properties. \\
We now highlight the main benefits of this design methodology.  The optimization of the emitted power amounts to optimize the absorptivity by the active medium. In practice, the design of a metasurface only requires a set of calculations of the absorbed power over the entire active region for a few angles and two polarization states. The technique thus requires less computation time than repeating the emission calculation by dipolar emitters averaging over all positions in a metasurface cell and over the three polarization states \cite{Vaskin19}.  Furthermore, it is important to stress that  Equation (1) gives an absolute expression of the emitted power. Hence, for a given chemical potential, it is seen that the emitted power has an upper bound corresponding to an absorptivity equal to 1. Kirchhoff's law thus enables to design a metasurface which approaches a maximum absorptivity. Previous design strategies aimed at optimizing the incident field in the active region owing to reciprocity but could not quantify if the design was close from the upper bound  \cite{Vaskin19,Liu18,Heki22}. In summary, Kirchhoff's law enables to design a directional source with maximum brightness.\\

We now turn to the implementation of this design methodolodgy. In order to perform absorption calculations by the active region, we model the layer of NPLs by an effective homogeneous medium with a complex refractive index. This approach is known to reproduce the reflectivity of a layer of NPLs on a metal surface. It also enables to capture the strong coupling between NPLs and plasmons \cite{Shlesinger19}. To do so, we do ellipsometry measurements to obtain the effective refractive index of the NPLs. The experimental data are given in the supplementary materials \cite{Supplementary}. An accurate refractive index model for the nanoplatelets in needed in order to model the photoluminescent emitted power with Kirchhoff's law, in particular at wavelengths far from the resonance \cite{Bailly22}. Ellipsometry data have the advantage of keeping things simple and it describes accurately the reflectivity measurements. It is thus a satisfactory model for the design of absorptivity. We also use an interpolation of the refractive index of silver of reference \cite{Palik12}, in agreement with the ellipsometry measurements (see \cite{Supplementary}). \\
To proceed, the absorptivity is computed with Rigorous Coupled-Wave Analysis (RCWA) \cite{Moharam81,Li97,Hugonin21} as a function of the angle of incidence at the peak emission wavelength of the NPLs (at 605 nm). A scan of the geometric parameters of the metasurface presented in Figure \ref{schema} is done until an absorptivity with a maximum at $\theta = 0^{\circ}$ is obtained, with a full-width at half-maximum (FWHM) smaller than 30$^{\circ}$. \\
We now provide a guideline to optimize the absorptivity in the NPLs in order to obtain the highest possible brightness. The parameters of the metasurface have been designed so that its total absorptivity is maximal. This can be achieved if the so-called critical coupling condition is satisfied \cite{Haus84,IntroNano22}, namely, if the radiative losses are equal to the ohmic losses in the materials. The losses of the materials are the sum of losses in the NPLs and in the metal. To enhance the absorptivity in the NPLs, we aimed at reducing the losses in the metal, and thus chose silver. As discussed in reference \cite{Paggi23}, the bare plasmonic grating must be in the overcoupled regime (radiative losses are larger than material losses) in order to ensure that material losses are equal to radiative losses when adding a layer of NPLs. 

\begin{figure}[h!]
\centering
\includegraphics[width = 0.5\textwidth]{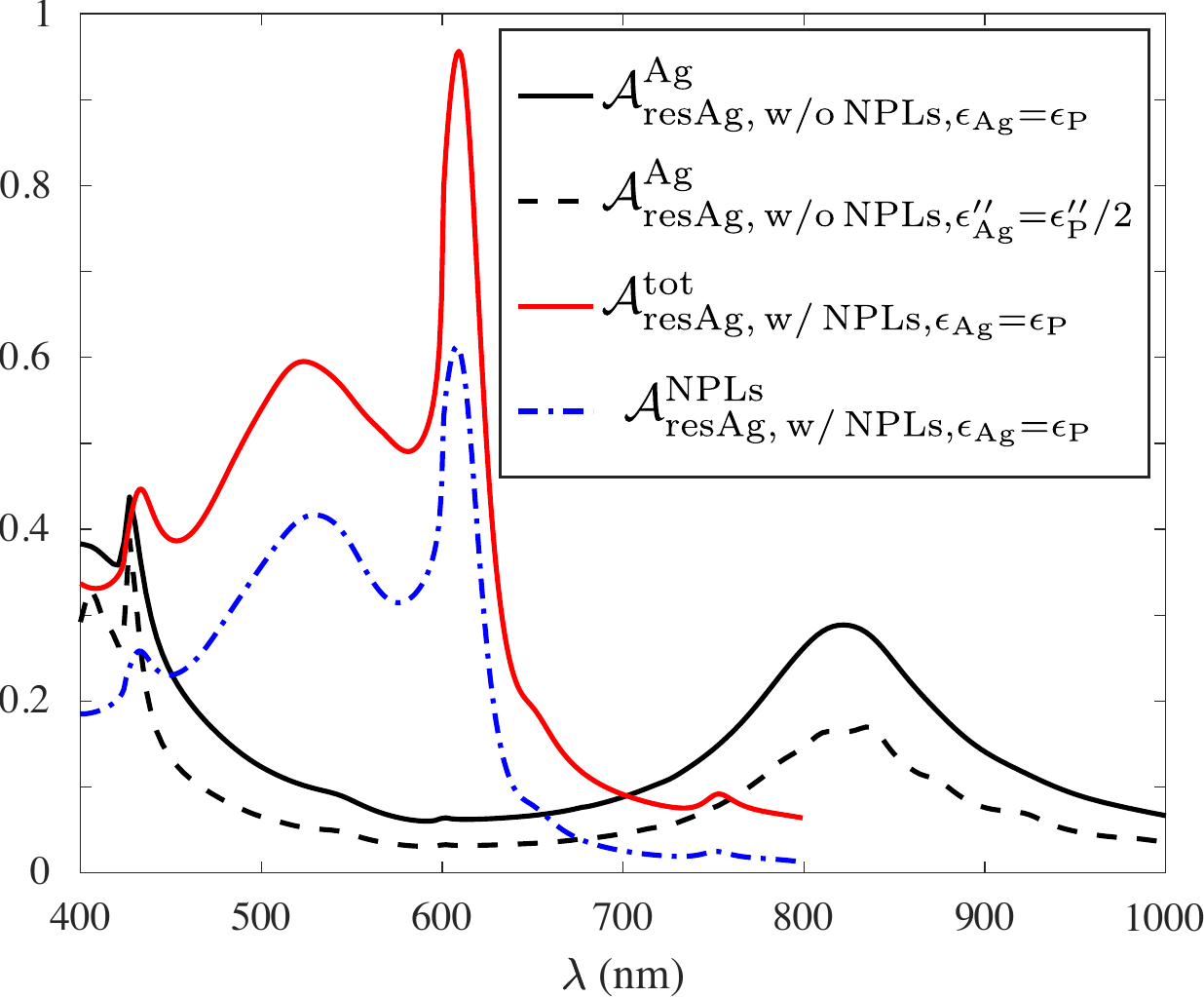}
\caption{Absorptivity as a function of the losses in silver, for a silver grating with or without the NPLs, where the geometric parameters of the grating are given in Figure \ref{schema}, at normal incidence. The dark solid line corresponds to the absorptivity in the bare silver grating, using the refractive index model of reference \cite{Palik12}, i.e. $\epsilon_{\mathrm{Ag}}=\epsilon'_{\mathrm{P}}+i\epsilon''_{\mathrm{P}}$. The dark dotted line corresponds to the absorptivity for the bare silver grating, when the losses in silver have been divided by a factor 2 i.e. $\epsilon''_{\mathrm{Ag}}=\epsilon''_{\mathrm{P}}/2$. The solid red line corresponds to the total absorptivity (in the silver and the NPLs) for the 2D grating with the NPLs. The blue dotted line corresponds to the absorptivity only in the NPLs for the 2D grating with the NPLs.}\label{couplage_critique}
\end{figure}

We plot the absorptivity at normal incidence in Figure \ref{couplage_critique} as a function of the wavelength (with and without the NPLs), in order to evidence the large absorptivity at critical coupling in the presence of NPLs. The solid red line corresponds to the total absorptivity (in silver and NPLs) in presence of the NPLs. At the peak emission wavelength of the NPLs, the absorptivity is close to 1. The value of absorptivity in the NPLs layer reaches 0.6 at 605 nm at normal incidence (see Figure \ref{couplage_critique}, blue dotted line). These results show that the absorptivity in the NPLs, and thus the brightness, have been optimized and reaches its highest possible value. 

To show that the metasurface is in the overcoupled regime, we plot the absorptivity of the bare silver grating in dark solid line with $\epsilon_{\mathrm{Ag}}=\epsilon'_{\mathrm{P}}+i\epsilon''_{\mathrm{P}}$ the silver refractive index. This curve shows a broad plasmonic resonance around 825 nm whose width is given by both radiative and non-radiative losses. 
To check that radiative losses are dominant, we reduce the losses in the metal by imposing  $\epsilon''_{\mathrm{Ag}}=\epsilon''_{\mathrm{P}}/2$. The associated absorptivity is plotted with the dark dotted line. The width of the resonance is barely modified, which proves that the resonator is over-coupled i.e. that the radiative losses dominate the losses and thus the resonance width. \\
Let us stress that the resonance of the metallic grating is much broader than the emitter spectral width. Hence, the same grating can be used for emitters at different wavelengths within the plasmon resonance width.\\

We have been able to obtain a bright and directional emission normal to the surface (see Equation \eqref{C1_eq_abs}). With this method, we obtain a trade-off between the highest possible value of absorptivity (i.e. emissivity) at $\theta = 0 ^{\circ}$ and the directionality. For instance, increasing the thickness of the NPLs on top of the grating may lead to a higher brightness but is detrimental to the directionality, a feature that could not be identified when designing the metasurface using the average emission of dipoles.
The parameters obtained by this procedure are equal to $h_{\mathrm{res}} = 100$ nm, $l_{\mathrm{res}} =$ 450 nm, $p_{\mathrm{res}} = $ 600 nm and $h_{\mathrm{top-NPLs}} =$ 2 nm. Initially, we considered that the NPLs fully filled the grooves without overfilling them, so that the simulations were run with $h_{\mathrm{top-NPLs}} =$ 0 nm. However, due to the experimental spin coating deposition of the nanoplatelets (see Methods), some nanoplatelets might have been left on top of the grating. We thus adjusted the value of $h_{\mathrm{top-NPLs}}$ and we obtained a better agreement with the experiment with $h_{\mathrm{top-NPLs}} = 2$ nm, which corresponds to a partially filled monolayer. The numerical computations with $h_{\mathrm{top-NPLs}} =$ 0 nm are given in the supplementary. 

This metasurface leads to a directional emission both in Transverse Electric (TE) and Transverse Magnetic (TM) polarizations and thus a directional total emission. Note that the design is performed using the total emitted power, adding the TE and TM polarizations. We show both polarizations in the supplementary \cite{Supplementary}. \\

\section{Results}\label{sec2}
\begin{figure}[h!]
\centering
\includegraphics[width = 0.4\textwidth]{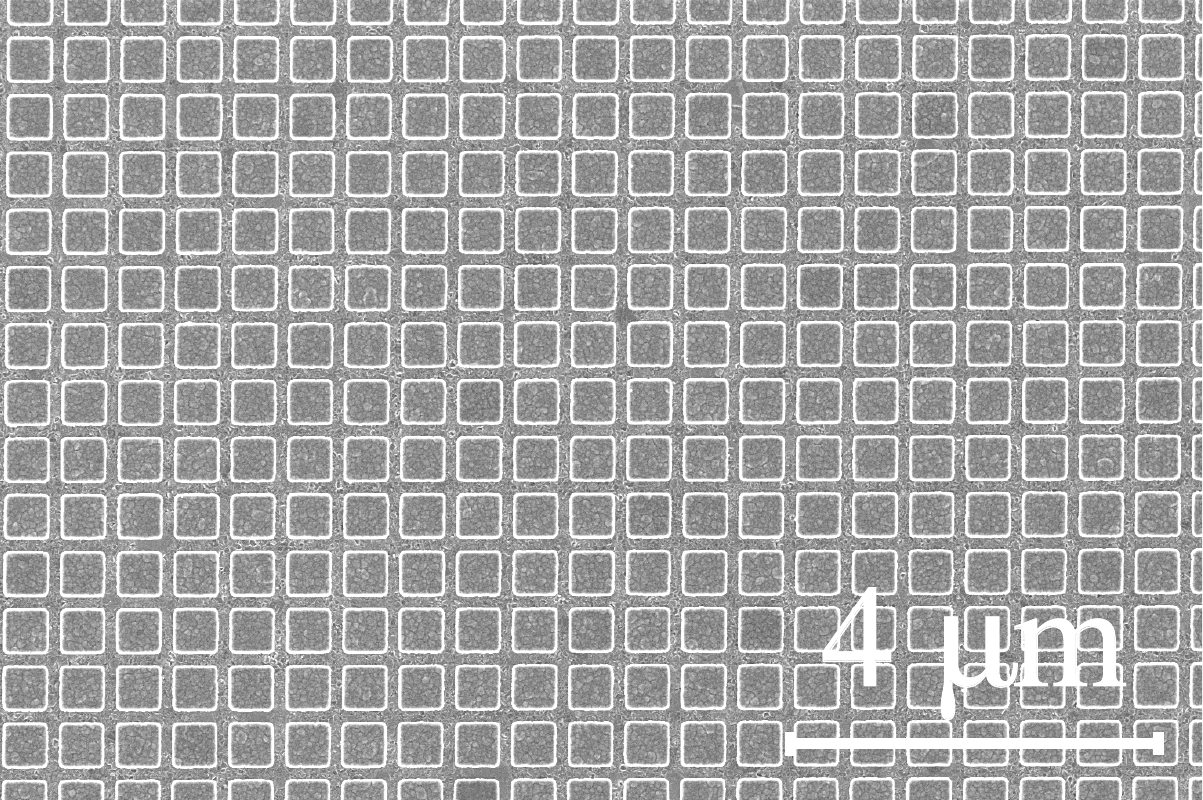}
\caption{Scanning electron microscopy (SEM) image of a 2D grating of silver, before nanoplatelets deposition.}\label{MEB}
\end{figure}

The sample was fabricated using electronic lithography on a silicon substrate and the layer of NPLs has been deposited by spin coating, as detailed in the Methods section. A SEM image of the grating before deposition of the nanoplatelets is presented in Figure \ref{MEB}. The width of the pillars has been evaluated from the SEM images, and it varies from 480 nm in the bottom to 450 nm in the top. In the numerical simulations we considered straight pillars with a uniform width of 450 nm. 

\begin{figure}[h!]
\centering
\includegraphics[width = 0.9\textwidth]{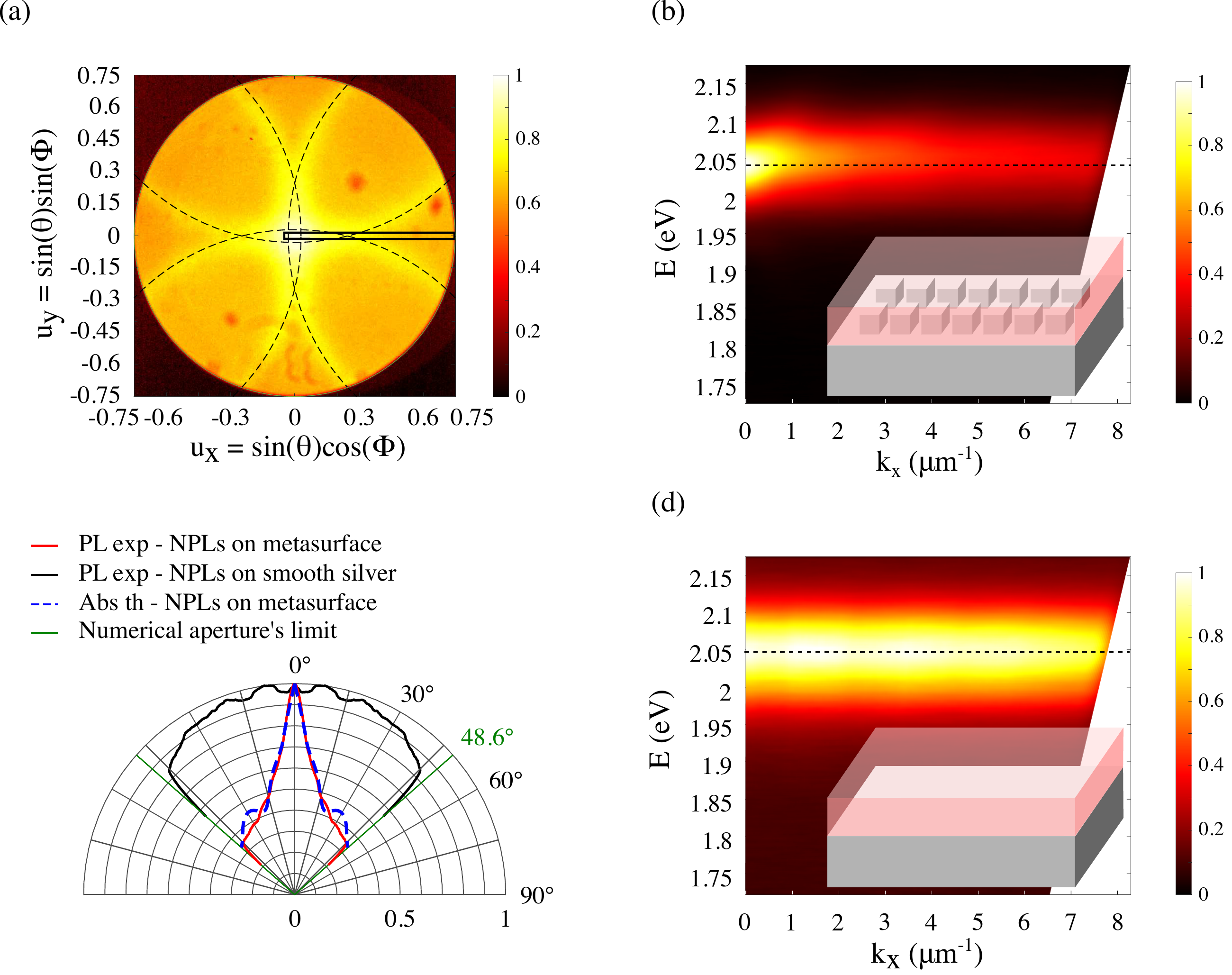}
\caption{Experimental radiation pattern. (a) Normalized Fourier image of the sample. The dotted lines corresponds to the theoretical isofrequency projection on the ($k_{x},k_{y}$) plane of the dispersion relation of a surface plasmon polariton existing in a 2 nm thick NPLs layer deposited on silver, after diffraction by a grating of $p_{\mathrm{res}} = 600$ nm, at 607.4 nm (see Figure \ref{directionality}c). The black rectangle corresponds to the slit position at the entrance of the spectrometer. (b) Photoluminescence intensity of a sample consisting of NPLs on a 2D metasurface as a function of the energy and the in-plane wave vector $k_{x}$ for NPLs on the 2D metasurface, (c) Normalized experimental radiation patterns for NPLs on the 2D metasurface (red curve) or on smooth silver (black line) at their experimental peak wavelength (in dotted lines in Figure b,d) and normalized theoretical absorption pattern (blue dotted line) calculated at the experimental peak emission wavelength (at 607.4 nm, black dotted line in Figure b). The green lines correspond to the numerical aperture's limit (NA = 0.75). (d) Photoluminescence as a function of the photon energy and the in-plane wave vector $k_{x}$ for NPLs on smooth silver. }\label{results_2D}
\end{figure}

The sample is pumped using the 436 nm line of a mercury vapor lamp. The light is focused on the sample by an objective with a numerical aperture of 0.75, probing an area whose diameter is around 110 \textmu m large. The photoluminescence is then collected through the same objective and the back focal plane of the microscope objective is imaged onto the slit of a spectrometer. A Fourier plane image as well as energy-momentum figures of photoluminescence from the optically pumped metasurfaces are shown in Figure \ref{results_2D}a,b. The emission is highly directional with a FWHM about $\pm 26.8^{\circ}$ at the peak wavelength (represented by the horizontal dotted line in Figure \ref{results_2D}b). 

An experimental comparison with a sample without the grating (that is with a flat, uniform and homogeneous emitter layer with no structuration) is given in Figure \ref{results_2D}c,d. In this latter case, the emission is quasi isotropic, as expected for an ensemble of point-like emitters (black line in Figure \ref{results_2D}c). The comparison between the experimental radiation pattern (red line in Figure \ref{results_2D}c) and the theoretical absorptivity pattern (blue dotted line in Figure \ref{results_2D}c) shows a good agreement between experiment and theory which validates the method of design. \\

\begin{figure}[h!]
\centering
\includegraphics[width = 1\textwidth]{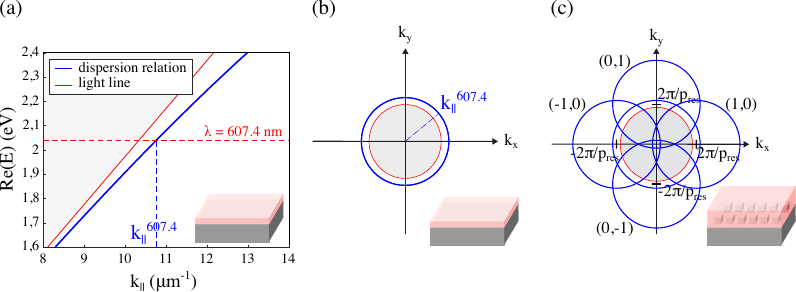}
\caption{(a) Dispersion relation of a surface plasmon at the interface between a silver substrate and a 2 nm thick layer of NPLs. (b) Schematic diagram of its isofrequency projection on the ($k_{x},k_{y}$) plane, at $\lambda = 607.4$ nm. (c) Schematic diagram of the isofrequency contour of the surface plasmon dispersion relation on the 2D metasurface with a periodicity $p_{\mathrm{res}}$.}\label{directionality}
\end{figure}

To interpret the far-field emission of Figure \ref{results_2D}a, we simply assume that the coupling is mediated by the surface plasmon propagating at the interface silver/NPLs/air. Indeed, the size of the pillars is much larger than the width of the thin grooves (Figure \ref{MEB}). Thus, the sample can be seen to first approximation as a translational invariant system, composed of NPLs on top of silver, which is perturbed by the presence of the grooves.
Therefore, we can plot the dispersion relation of the surface plasmons propagating at the interface between a 2 nm thick layer of NPLs and a silver substrate, see Figure \ref{directionality}a. It has been obtained by searching numerically the poles of the reflectivity of the system, with a complex frequency (or energy E (eV)) and a real wavevector $k_{\parallel}$ \cite{Archambault09,benisty_greffet_lalanne}. To do so, we fitted the refractive indexes of the NPLs and silver by a polynome (see \cite{Supplementary}). At the experimental peak emission wavelength (607.4 nm), the wavevector is $k_{\parallel}^{607.4} = 10.78$ \textmu$\mathrm{m}{^{-1}}$. In the plane $(k_x,k_y)$, the corresponding dispersion relation is a circle (Figure \ref{directionality}b). 

We now take into account the two-dimensional periodicity of the metasurface, by periodizing the initial dispersion relation in the momentum plane in both $k_x$ and $k_y$ direction (see Figure \ref{directionality}c). If we superimpose the circles of Figure \ref{directionality}c on the Fourier image of Figure \ref{results_2D}a (normalizing the radius and the circles' center's positions by $k_{0} = 2\pi/\lambda$, with $\lambda = 607.4$ nm) we show that we can reproduce accurately the observed emission pattern along the arcs of circles. In summary, the emission is enhanced by coupling to leaky surface plasmons. The choice of the periodicity is critical to ensure a peak at normal incidence. \\

To characterize the directionality of the source, we use a figure of merit called "beam efficiency", defined in reference \cite{Iyer20} as  the fraction of total light intensity collected within one FWMH along the $k_{x}$ axis at the peak emission frequency. Since the collected light is limited by the numerical aperture, we cannot measure the total emitted intensity. We derive an upper bound and a lower bound by assuming that the emitted power into angles larger than $45.3^{\circ}$ is either null or equal to its value at $45.3^{\circ}$. We obtain an efficiency in the range 35$\%$ and 44.8 $\%$ , larger than previous reports (see Supplementary \cite{Supplementary}).\\
Note that the enhancement due to the metasurface enables to achieve large brightness with a small amount of active material. At the experimental peak emission wavelength, the absorptivity is 0.61. To reach the same absorptivity with a nanoplatelets layer on a glass substrate at normal incidence, the thickness layer should be 850 nm, corresponding to a volume 18.6 times larger.

\section{Conclusion}
In summary, we have shown that the extension of Kirchhoff's law to photoluminescent metasurfaces provides an accurate model of the emission. To our knowledge, it is the first time that this law has been used to design a photoluminescent metasurface with specific emission properties. Here, we have shown that it is possible to design a bright metasurface emitting within a narrow spectral range and within a narrow cone with half-angle below 15°, taking advantage of surface plasmons. Such a device has been realized for the first time with nanoplatelets in the visible range. We have obtained a beam efficiency on the order of $40\%$ as well as an absorptivity of 0.61. The same approach could be used to design photoluminescent metasurfaces controlling the emitted polarization as reported recently in the context of thermal emission \cite{Nguyen23}. The method could also be applied to optimize dielectric photoluminescent metasurfaces \cite{Liu18,Decker16,Bucher19} which could enable sources with narrower emission cones, as well as electroluminescent metasurfaces \cite{LeVan,Bossavit23,Dabard23}. 


\section{Methods}

\threesubsection{Lift-off process for the silver metasurface}\\
The silver grating was fabricated by lift-off process. First, a 1-nm-thick germanium layer was deposited on a silicon substrate in order to facilitate the wetting on the substrate, followed by a 200-nm-thick layer of silver deposition, by electron beam evaporation (at 0.1 nm/s). A PMMA A4 e-beam resist was then spin coated on the sample (4000 rpm during 60 seconds with an acceleration of 2000 rpm/s) and baked at 180 $^\circ$C for 2 minutes. E-beam lithography was processed at 100 keV with a dose of 800 \textmu C/cm2 and a 5 nA current on a Raith EBPG5200 machine. The sample was then developed in MIBK:IPA (1:3) for 1 minute at 20$^{\circ}$, rinsed with IPA (for 30 seconds) and dried with nitrogen. A 100-nm-thick layer of silver was then deposited by electron beam evaporation (at 0.1 nm/s) and the final 600 × 600 \textmu m$^2$ structure was obtained by immersing the sample the whole night in butanone. The sample was then rinsed with IPA and dried with nitrogen.\\

\threesubsection{Spin coating of nanoplatelets}\\
The nanoplatelet solution was fabricated by Corentin Dabard and Sandrine Ithurria (Laboratoire de Physique et d’Etude des Mat{\'e}riaux, PSL Research University, CNRS UMR 8213, UPMC Sorbonne Universit{\'e}, ESPCI Paris, 10 rue Vauquelin, 75005 Paris, France). The nanoplatelets were dispersed in Toluene (with a concentration C $\approx 3.8$ mg.mL$^{-1}$). 200 \textmu L of nanoplatelet solution was spincoated on the metallic grating at 500 rpm during 30 seconds with an acceleration ramp of 5 seconds. When the nanoplatelets are deposited on top of a flat silver substrate with these spin coating parameters, it leads to a layer of thickness $42\pm5$ nm thick. The thickness has been measured by doing a scratch on the sample and measuring the height of the slit by AFM \cite{Supplementary}. When depositing the nanoplatelets on the grating sample, the nanoplatelets fill the grooves since the walls of the grating are much higher than the thickness of the nanoplatelets layer on a flat substrate. We considered that very few nanoplatelets remained on top of the grating, allowing in our simulation the parameter $h_{\mathrm{top-NPLs}}$ to be non-zero and leading to a value of 2 nm to optimize the comparison with data. \\

\medskip
\textbf{Supporting Information} \par 

Supporting Information is available.

\medskip
\textbf{Acknowledgements} \par 
This work is supported by the French National Agency (ANR) (ANR-17-CE24-0046). J.-J.G. acknowledges the support of Institut Universitaire de France (IUF). SIL acknowledges the funding from the European Research Council (ERC) under the European Union’s Horizon 2020 research and innovation programme (Ne2DeM Grant Agreement No. 853049). We thank Xavier Lafosse from Centre de Nanosciences et de Nanotechnologie, Universit\'e Paris-Saclay, CNRS, 91120 Palaiseau, for the ellipsometry measurements. 
\medskip

%


\begin{thebibliography}{1}

\bibitem{Palik12}
E.~Palik, {\em Handbook of Optical Constants of Solids: Volume 1}.
\newblock No.~vol.~1, Elsevier Science, 2012.

\bibitem{article}
 See Article.

\end{thebibliography}


\begin{thebibliography}{10}

\bibitem{Li18}
A.~Li, S.~Singh, and D.~Sievenpiper, ``Metasurfaces and their applications,''
  {\em Nanophotonics}, vol.~7, no.~6, pp.~989--1011, 2018.

\bibitem{Yu14}
N.~Yu and F.~Capasso, ``Flat optics with designer metasurfaces,'' {\em Nature
  Materials}, vol.~13, no.~2, pp.~139--150, 2014.

\bibitem{Yu11}
N.~Yu, P.~Genevet, M.~A. Kats, F.~Aieta, J.-P. Tetienne, F.~Capasso, and
  Z.~Gaburro, ``Light propagation with phase discontinuities: Generalized laws
  of reflection and refraction,'' {\em Science}, vol.~334, no.~6054,
  pp.~333--337, 2011.

\bibitem{Vecchi09}
G.~Vecchi, V.~Giannini, and J.~G{\'{o}}mez~Rivas, ``Shaping the fluorescent
  emission by lattice resonances in plasmonic crystals of nanoantennas,'' {\em
  Phys. Rev. Lett.}, vol.~102, p.~146807, Apr 2009.

\bibitem{Steele10}
J.~M. Steele, I.~Gagnidze, and S.~M. Wiele, ``Efficient extraction of
  fluorescence emission utilizing multiple surface plasmon modes from gold wire
  gratings,'' {\em Plasmonics}, vol.~5, no.~3, pp.~319--324, 2010.

\bibitem{Lozano13}
G.~Lozano, D.~J. Louwers, S.~Rodríguez, S.~Murai, O.~T.~A. Jansen, M.~A.
  Verschuuren, and J.~G\'{o}mez~Rivas, ``Plasmonics for solid-state lighting:
  enhanced excitation and directional emission of highly efficient light
  sources,'' {\em Light: Science \& Applications}, vol.~2, p.~e66, 2013.

\bibitem{Lozano14}
G.~Lozano, G.~Grzela, M.~A. Verschuuren, M.~Ramezani, and J.~G. Rivas,
  ``Tailor-made directional emission in nanoimprinted plasmonic-based
  light-emitting devices,'' {\em Nanoscale}, vol.~6, pp.~9223--9229, 2014.

\bibitem{WangS18}
S.~Wang, Q.~Le-Van, T.~Peyronel, M.~Ramezani, N.~Van~Hoof, T.~G. Tiecke, and
  J.~Gómez~Rivas, ``Plasmonic nanoantenna arrays as efficient etendue reducers
  for optical detection,'' {\em ACS Photonics}, vol.~5, no.~6, pp.~2478--2485,
  2018.

\bibitem{Vaskin18}
A.~Vaskin, J.~Bohn, K.~E. Chong, T.~Bucher, M.~Zilk, D.-Y. Choi, D.~N. Neshev,
  Y.~S. Kivshar, T.~Pertsch, and I.~Staude, ``Directional and spectral shaping
  of light emission with mie-resonant silicon nanoantenna arrays,'' {\em ACS
  Photonics}, vol.~5, no.~4, pp.~1359--1364, 2018.

\bibitem{DiMaria13}
J.~DiMaria, E.~Dimakis, T.~D. Moustakas, and R.~Paiella, ``{Plasmonic off-axis
  unidirectional beaming of quantum-well luminescence},'' {\em Applied Physics
  Letters}, vol.~103, p.~251108, 12 2013.

\bibitem{Iyer20}
P.~P. Iyer, R.~A. DeCrescent, Y.~Mohtashami, G.~Lheureux, N.~A. Butakov,
  A.~Alhassan, C.~Weisbuch, S.~Nakamura, S.~P. DenBaars, and J.~A. Schuller
  {\em Nature Photonics}, vol.~14, no.~9, pp.~543--548, 2020.

\bibitem{Heki22}
L.~Heki, Y.~Mohtashami, R.~A. DeCrescent, A.~Alhassan, S.~Nakamura, S.~P.
  DenBaars, and J.~A. Schuller, ``Designing highly directional luminescent
  phased-array metasurfaces with reciprocity-based simulations,'' {\em ACS
  Omega}, vol.~7, no.~26, pp.~22477--22483, 2022.

\bibitem{Rodriguez12}
S.~R.~K. Rodriguez, G.~Lozano, M.~A. Verschuuren, R.~G{\'{o}}mez, K.~Lambert,
  B.~De~Geyter, A.~Hassinen, D.~Van~Thourhout, Z.~Hens, and
  J.~G{\'{o}}mez~Rivas, ``Quantum rod emission coupled to plasmonic lattice
  resonances: A collective directional source of polarized light,'' {\em
  Applied Physics Letters}, vol.~100, no.~11, p.~111103, 2012.

\bibitem{GuoTorma15}
R.~Guo, S.~Derom, A.~I. V\"{a}kev\"{a}inen, R.~J.~A. van Dijk-Moes,
  P.~Liljeroth, D.~Vanmaekelbergh, and P.~T\"{o}rm\"{a}, ``Controlling quantum
  dot emission by plasmonic nanoarrays,'' {\em Opt. Express}, vol.~23,
  pp.~28206--28215, Nov 2015.

\bibitem{WangX20}
X.~Wang, Y.~Li, R.~Toufanian, L.~C. Kogos, A.~M. Dennis, and R.~Paiella,
  ``Geometrically tunable beamed light emission from a quantum-dot ensemble
  near a gradient metasurface,'' {\em Advanced Optical Materials}, vol.~8,
  no.~8, p.~1901951, 2020.

\bibitem{Dabard23}
C.~Dabard, E.~Bossavit, T.~H. Dang, N.~Ledos, M.~Cavallo, A.~Khalili, H.~Zhang,
  R.~Alchaar, G.~Patriarche, A.~Vasanelli, B.~T. Diroll, A.~Degiron,
  E.~Lhuillier, and S.~Ithurria, ``Electroluminescence and plasmon-assisted
  directional photoluminescence from 2d hgte nanoplatelets,'' {\em The Journal
  of Physical Chemistry C}, vol.~127, no.~30, pp.~14847--14855, 2023.

\bibitem{Bossavit23}
E.~Bossavit, T.~H. Dang, P.~He, M.~Cavallo, A.~Khalili, C.~Dabard, H.~Zhang,
  D.~Gacemi, M.~G. Silly, C.~Abadie, B.~Gallas, D.~Pierucci, Y.~Todorov,
  C.~Sirtori, B.~T. Diroll, A.~Degiron, E.~Lhuillier, and A.~Vasanelli,
  ``Plasmon-assisted directional infrared photoluminescence of hgte
  nanocrystals,'' {\em Advanced Optical Materials}, vol.~n/a, no.~n/a,
  p.~2300863, 2023.

\bibitem{Greffet02}
J.-J. Greffet, R.~Carminati, K.~Joulain, J.-P. Mulet, S.~Mainguy, and Y.~Chen,
  ``Coherent emission of light by thermal sources,'' {\em Nature}, vol.~416,
  p.~61, 2002.

\bibitem{Vaskin19}
A.~Vaskin, R.~Kolkowski, A.~F. Koenderink, and I.~Staude, ``Light-emitting
  metasurfaces,'' {\em Nanophotonics}, vol.~8, no.~7, pp.~1151 -- 1198, 2019.

\bibitem{Lozano16}
G.~Lozano, S.~R.~K. Rodriguez, M.~A. Verschuuren, and J.~G{\'{o}}mez~Rivas,
  ``{Metallic nanostructures for efficient LED lighting},'' {\em Light: Science
  {\&} Applications}, vol.~5, no.~6, pp.~e16080--e16080, 2016.

\bibitem{Liu18}
S.~Liu, A.~Vaskin, S.~Addamane, B.~Leung, M.-C. Tsai, Y.~Yang, P.~P.
  Vabishchevich, G.~A. Keeler, G.~Wang, X.~He, Y.~Kim, N.~F. Hartmann,
  H.~Htoon, S.~K. Doorn, M.~Zilk, T.~Pertsch, G.~Balakrishnan, M.~B. Sinclair,
  I.~Staude, and I.~Brener, ``Light-emitting metasurfaces: Simultaneous control
  of spontaneous emission and far-field radiation,'' {\em Nano Letters},
  vol.~18, no.~11, pp.~6906--6914, 2018.
\newblock PMID: 30339762.

\bibitem{Costantini15}
D.~Costantini, A.~Lefebvre, A.-L. Coutrot, I.~Moldovan-Doyen, J.-P. Hugonin,
  S.~Boutami, F.~Marquier, H.~Benisty, and J.-J. Greffet, ``Plasmonic
  metasurface for directional and frequency-selective thermal emission,'' {\em
  Phys. Rev. Appl.}, vol.~4, p.~014023, Jul 2015.

\bibitem{Wojszvzyk21}
L.~Wojszvzyk, A.~Nguyen, A.-L. Coutrot, C.~Zhang, B.~Vest, and J.-J. Greffet,
  ``An incandescent metasurface for quasimonochromatic polarized mid-wave
  infrared emission modulated beyond 10 mhz,'' {\em Nature Communications},
  vol.~12, no.~1, p.~1492, 2021.

\bibitem{Nguyen22}
A.~Nguyen and J.-J. Greffet, ``Efficiency optimization of mid-infrared
  incandescent sources with time-varying temperature,'' {\em Opt. Mater.
  Express}, vol.~12, pp.~225--239, Jan 2022.

\bibitem{Nguyen23}
A.~Nguyen, J.-P. Hugonin, A.-L. Coutrot, E.~Garcia-Caurel, B.~Vest, and J.-J.
  Greffet, ``Large circular dichroism in the emission from an incandescent
  metasurface,'' {\em Optica}, vol.~10, pp.~232--238, Feb 2023.

\bibitem{Watts12}
C.~M. Watts, X.~Liu, and W.~J. Padilla, ``Metamaterial electromagnetic wave
  absorbers,'' {\em Advanced Materials}, vol.~24, no.~23, pp.~OP98--OP120,
  2012.

\bibitem{Baranov19}
D.~G. Baranov, Y.~Xiao, I.~A. Nechepurenko, A.~Krasnok, A.~Alù, and M.~A.
  Kats, ``Nanophotonic engineering of far-field thermal emitters,'' {\em Nature
  Materials}, vol.~18, no.~9, pp.~920--930, 2019.

\bibitem{Overvig21}
A.~C. Overvig, S.~A. Mann, and A.~Al\`u, ``Thermal metasurfaces: Complete
  emission control by combining local and nonlocal light-matter interactions,''
  {\em Phys. Rev. X}, vol.~11, p.~021050, Jun 2021.

\bibitem{Arnold12}
C.~Arnold, F.~Marquier, M.~Garin, F.~Pardo, S.~Collin, N.~Bardou, J.-L.
  Pelouard, and J.-J. Greffet, ``Coherent thermal infrared emission by
  two-dimensional silicon carbide gratings,'' {\em Phys. Rev. B}, vol.~86,
  p.~035316, Jul 2012.

\bibitem{WangX23}
X.~Wang, T.~Sentz, S.~Bharadwaj, S.~K. Ray, Y.~Wang, D.~Jiao, L.~Qi, and
  Z.~Jacob, ``Observation of nonvanishing optical helicity in thermal radiation
  from symmetry-broken metasurfaces,'' {\em Science Advances}, vol.~9, no.~4,
  p.~eade4203, 2023.

\bibitem{Monin23}
H.~Monin, A.~Loirette-Pelous, E.~De~Leo, A.~A. Rossinelli, F.~Prins, D.~J.
  Norris, E.~Bailly, J.-P. Hugonin, B.~Vest, and J.-J. Greffet, ``Controlling
  light emission by a thermalized ensemble of colloidal quantum dots with a
  metasurface,'' {\em Opt. Express}, vol.~31, pp.~4851--4861, Jan 2023.

\bibitem{Bailly22}
E.~Bailly, K.~Chevrier, C.~Perez de~la Vega, J.-P. Hugonin, Y.~De~Wilde,
  V.~Krachmalnicoff, B.~Vest, and J.-J. Greffet, ``Method to measure the
  refractive index for photoluminescence modelling,'' {\em Opt. Mater.
  Express}, vol.~12, pp.~2772--2781, Jul 2022.

\bibitem{Perez23}
C.~R. Pérez de~la Vega, E.~Bailly, K.~Chevrier, B.~Vest, J.-P. Hugonin,
  A.~Bard, A.~Gassenq, C.~Symonds, J.-M. Benoit, J.~Bellessa, J.-J. Greffet,
  Y.~De~Wilde, and V.~Krachmalnicoff, ``Plasmon-mediated energy transfer
  between two systems out of equilibrium,'' {\em ACS Photonics}, vol.~10,
  no.~4, pp.~1169--1176, 2023.

\bibitem{Caillas21}
A.~Caillas, S.~Suffit, P.~Filloux, E.~Lhuillier, and A.~Degiron,
  ``Identification of two regimes of carrier thermalization in pbs nanocrystal
  assemblies,'' {\em The Journal of Physical Chemistry Letters}, vol.~12,
  no.~21, pp.~5123--5131, 2021.
\newblock PMID: 34029086.

\bibitem{Greffet18}
J.-J. Greffet, P.~Bouchon, G.~Brucoli, and F.~Marquier, ``Light emission by
  nonequilibrium bodies: local kirchhoff law,'' {\em Physical Review X},
  vol.~8, no.~2, p.~021008, 2018.

\bibitem{Bailly21}
E.~Bailly, J.-P. Hugonin, B.~Vest, and J.-J. Greffet, ``Spatial coherence of
  light emitted by thermalized ensembles of emitters coupled to surface
  waves,'' {\em Phys. Rev. Res.}, vol.~3, p.~L032040, Aug 2021.

\bibitem{Diroll23}
B.~T. Diroll, B.~Guzelturk, H.~Po, C.~Dabard, N.~Fu, L.~Makke, E.~Lhuillier,
  and S.~Ithurria, ``2d ii–vi semiconductor nanoplatelets: From material
  synthesis to optoelectronic integration,'' {\em Chemical Reviews}, vol.~123,
  no.~7, pp.~3543--3624, 2023.
\newblock PMID: 36724544.

\bibitem{Altintas19}
Y.~Altintas, U.~Quliyeva, K.~Gungor, O.~Erdem, Y.~Kelestemur, E.~Mutlugun,
  M.~V. Kovalenko, and H.~V. Demir, ``Highly stable, near-unity efficiency
  atomically flat semiconductor nanocrystals of cdse/zns hetero-nanoplatelets
  enabled by zns-shell hot-injection growth,'' {\em Small}, vol.~15, no.~8,
  p.~1804854, 2019.

\bibitem{Rossinelli17}
A.~A. Rossinelli, A.~Riedinger, P.~Marqués-Gallego, P.~N. Knüsel, F.~V.
  Antolinez, and D.~J. Norris, ``High-temperature growth of thick-shell
  cdse/cds core/shell nanoplatelets,'' {\em Chem. Commun.}, vol.~53,
  pp.~9938--9941, 2017.

\bibitem{Baruj23}
H.~D. Baruj, I.~Bozkaya, B.~Canimkurbey, A.~T. Isik, F.~Shabani, S.~Delikanli,
  S.~Shendre, O.~Erdem, F.~Isik, and H.~V. Demir, ``Highly-directional,
  highly-efficient solution-processed light-emitting diodes of all-face-down
  oriented colloidal quantum well self-assembly,'' {\em Small}, vol.~19,
  no.~29, p.~2206582, 2023.

\bibitem{Supplementary}
 See Supplemental Material.

\bibitem{Rytov89}
S.~M. Rytov, Y.~A. Kravtsov, and V.~I. Tatarskii, {\em Principles of
  statistical radiophysics. 3. Elements of random fields}, vol.~3.
\newblock Berlin: Springer, 1989.

\bibitem{Wurfel82}
P.~Wurfel, ``The chemical potential of radiation,'' {\em Journal of Physics C:
  Solid State Physics}, vol.~15, no.~18, p.~3967, 1982.

\bibitem{WangH18}
H.~Wang, A.~Aassime, X.~Le~Roux, N.~J. Schilder, J.-J. Greffet, and A.~Degiron,
  ``Revisiting the role of metallic antennas to control light emission by lead
  salt nanocrystal assemblies,'' {\em Physical Review Applied}, vol.~10, no.~3,
  p.~034042, 2018.

\bibitem{Shlesinger19}
I.~Shlesinger, H.~Monin, J.~Moreau, J.-P. Hugonin, M.~Dufour, S.~Ithurria,
  B.~Vest, and J.-J. Greffet, ``{Strong Coupling of Nanoplatelets and Surface
  Plasmons on a Gold Surface},'' {\em ACS Photonics}, vol.~6, no.~11,
  pp.~2643--2648, 2019.

\bibitem{Palik12}
E.~Palik, {\em Handbook of Optical Constants of Solids: Volume 1}.
\newblock No.~vol.~1, Elsevier Science, 2012.

\bibitem{Moharam81}
M.~G. Moharam and T.~K. Gaylord, ``Rigorous coupled-wave analysis of
  planar-grating diffraction,'' {\em J. Opt. Soc. Am.}, vol.~71, pp.~811--818,
  Jul 1981.

\bibitem{Li97}
L.~Li, ``New formulation of the fourier modal method for crossed surface-relief
  gratings,'' {\em J. Opt. Soc. Am. A}, vol.~14, pp.~2758--2767, Oct 1997.

\bibitem{Hugonin21}
J.-P. Hugonin and P.~Lalanne, ``Reticolo software for grating analysis,'' 2021.

\bibitem{Haus84}
H.~Haus, {\em Waves and Fields in Optoelectronics}.
\newblock Prentice-Hall series in solid state physical electronics,
  Prentice-Hall, 1984.

\bibitem{IntroNano22}
H.~Benisty, J.~Greffet, and P.~Lalanne, {\em Introduction to Nanophotonics}.
\newblock Oxford Graduate Texts, Oxford University Press, 2022.

\bibitem{Paggi23}
L.~Paggi, A.~Fabas, H.~El~Ouazzani, J.-P. Hugonin, N.~Fayard, N.~Bardou,
  C.~Dupuis, J.-J. Greffet, and P.~Bouchon, ``Over-coupled resonator for
  broadband surface enhanced infrared absorption (seira),'' {\em Nature
  Communications}, vol.~14, no.~1, p.~4814, 2023.

\bibitem{Archambault09}
A.~Archambault, T.~V. Teperik, F.~Marquier, and J.-J. Greffet, ``Surface
  plasmon fourier optics,'' {\em Phys. Rev. B}, vol.~79, p.~195414, May 2009.

\bibitem{benisty_greffet_lalanne}
H.~Benisty, J.-J. Greffet, and P.~Lalanne, {\em Introduction to Nanophotonics}.
\newblock Oxford Graduate Texts, Oxford University Press, 2022.

\bibitem{Decker16}
M.~Decker and I.~Staude, ``Resonant dielectric nanostructures: a low-loss
  platform for functional nanophotonics,'' {\em Journal of Optics}, vol.~18,
  p.~103001, sep 2016.

\bibitem{Bucher19}
T.~Bucher, A.~Vaskin, R.~Mupparapu, F.~J.~F. L\"{o}chner, A.~George, K.~E.
  Chong, S.~Fasold, C.~Neumann, D.-Y. Choi, F.~Eilenberger, F.~Setzpfandt,
  Y.~S. Kivshar, T.~Pertsch, A.~Turchanin, and I.~Staude, ``Tailoring
  photoluminescence from mos2 monolayers by mie-resonant metasurfaces,'' {\em
  ACS Photonics}, vol.~6, no.~4, pp.~1002--1009, 2019.

\bibitem{LeVan}
Q.~Le-Van, X.~Le~Roux, A.~Aassime, and A.~Degiron, ``Electrically driven
  optical metamaterials,'' {\em Nature Communications}, vol.~7, no.~12017,
  pp.~2041--1723, 2016.

\end{thebibliography}




\end{document}



\title{2D silver-nanoplatelets metasurface for bright
directional photoluminescence designed with the
local Kirchhoff’s law: Supplemental material}

\maketitle


\author{Elise Bailly}
\author{Jean-Paul Hugonin}
\author{Jean-René Coudevylle}
\author{Corentin Dabard}
\author{Sandrine Ithurria}
\author{Benjamin Vest}
\author{Jean-Jacques Greffet*}

\begin{affiliations}
E. Bailly, J.-P. Hugonin, B. Vest, J.-J. Greffet\\
Universit\'e Paris-Saclay, Institut d’Optique Graduate School, CNRS, Laboratoire Charles Fabry, 91120 Palaiseau, France\\
Email Address: jean-jacques.greffet@institutoptique.fr

J.-R. Coudevylle\\
Centre de Nanosciences et de Nanotechnologies, Universit\'e Paris-Saclay, CNRS, 91120 Palaiseau, France

C. Dabard, S. Ithurria\\
Laboratoire de Physique et d’Etude des Mat\'eriaux, ESPCI-Paris, PSL Research University, Sorbonne Universit\'e UPMC Univ Paris
06, CNRS, 10 Rue Vauquelin, 75005 Paris, France

\end{affiliations}


\keywords{metasurfaces, photoluminescence, directionality, Kirchhoff}

\section{Nanoplatelets' properties}
\subsection{Emission and Absorption spectra of the nanoplatelets}
\begin{figure}[h!]
\centering
\includegraphics[width = 0.5\textwidth]{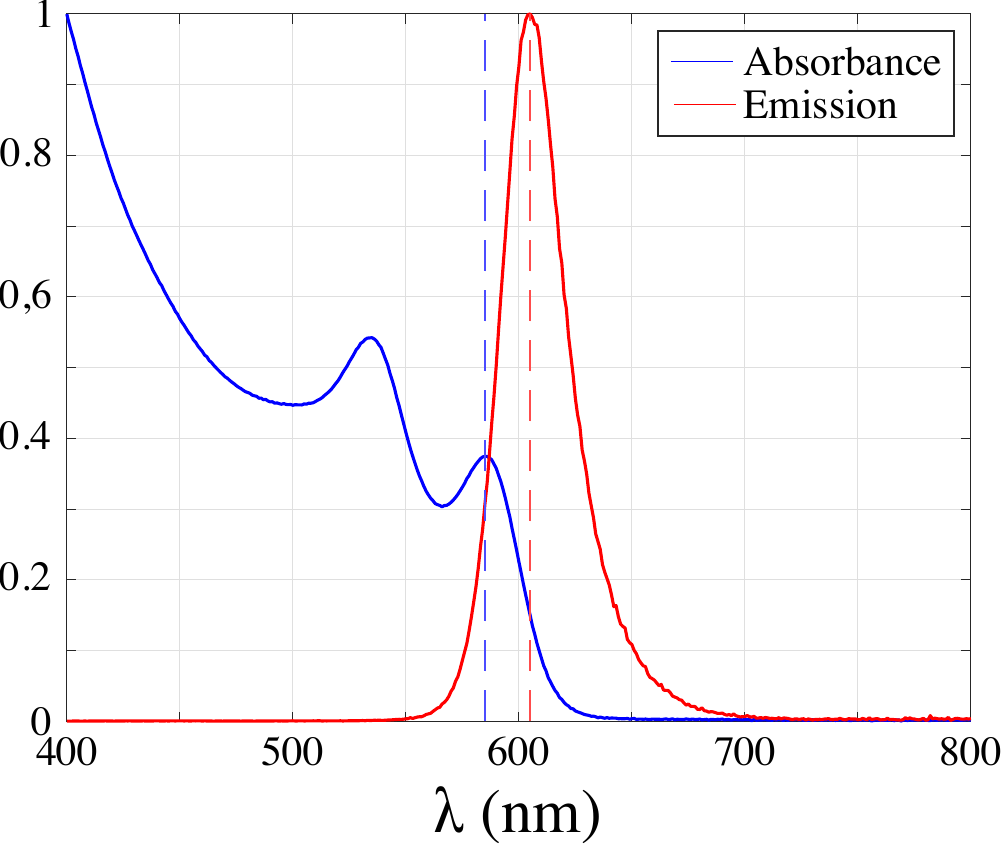}
\caption{Normalized emission and absorption spectra of the solution of NPLs in hexane in arbitrary units.}\label{fig_spectra}
\end{figure}

The absorption and emission spectra of the nanoplatelets (NPLs) (in solution in hexane) are given in Figure \ref{fig_spectra}. The peak emission wavelength is 605 nm. The Stokes shift, evaluated between the absorption peak (blue dotted line) and the emission peak (red dotted line) is 20 nm. 

\newpage

\subsection{TEM images}
The TEM (transmission electron microscopy) image of the NPLs is presented in Figure \ref{fig_TEM}. For TEM imaging, a drop of diluted NPLs solution in hexane is drop-casted on a copper grid covered with an amorphous carbon film. The grid is degassed overnight under secondary vacuum. A JEOL 2010F is used at 200 kV for the picture acquisition.

\begin{figure}[h!]
\centering
\includegraphics[width = 0.4\textwidth]{fig_supp_TEM_NPLs.pdf}
\caption{TEM image of the NPLs.}\label{fig_TEM}
\end{figure}

\subsection{Refractive index of the nanoplatelets}

\begin{figure}[h!]
\centering
\includegraphics[width = 0.4\textwidth]{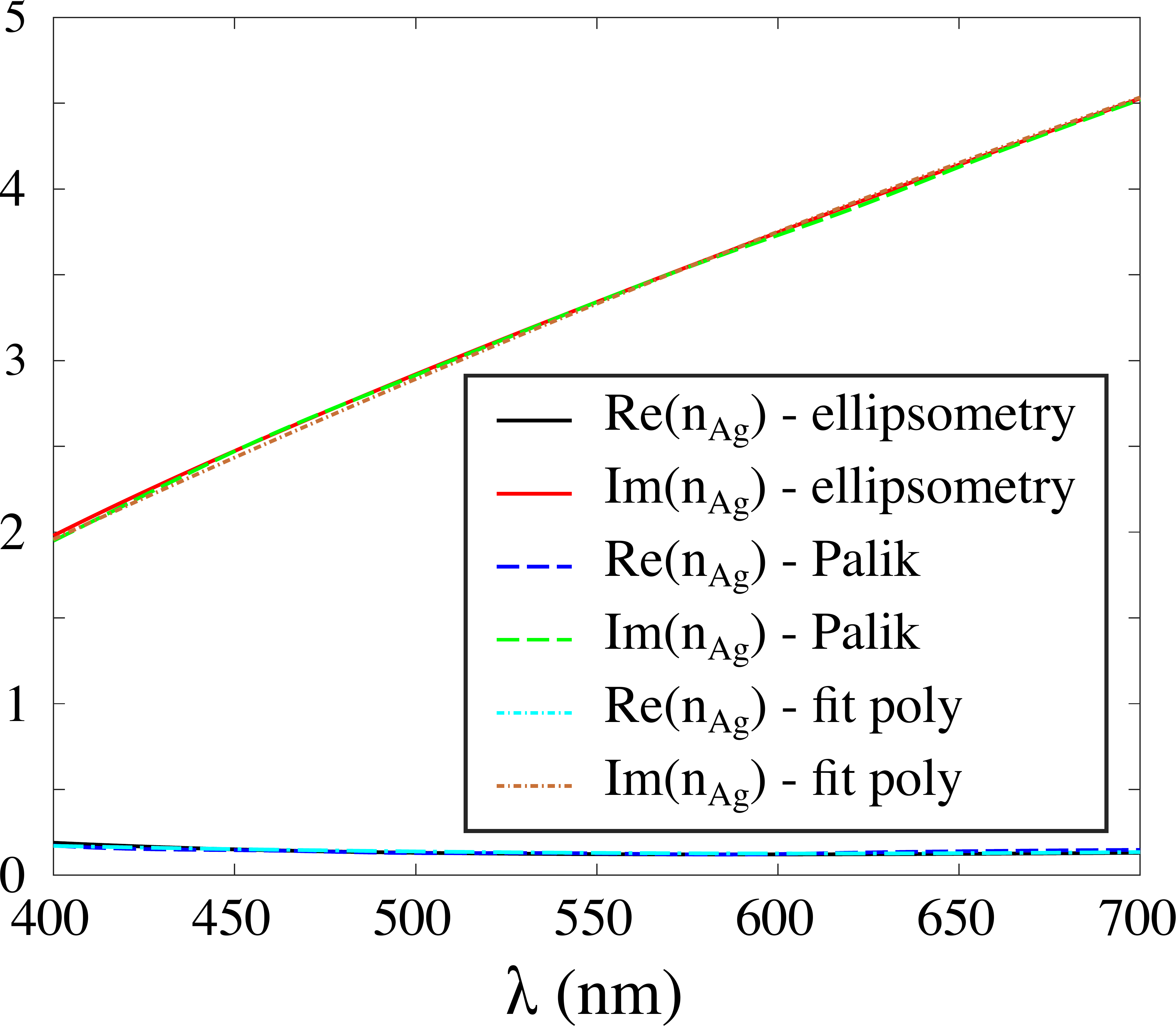}
\caption{Refractive index of silver, measured by ellipsometry from a sample composed of 50 nm of silver on top of a 1 nm of Germanium on a SF10 glass substrate. The experimental values are compared with the refractive index of silver of reference \cite{Palik12}, named "Palik" and the polynomial fit whose coefficients are given in Table \ref{tab_coeff_indice}.}\label{indice_Ag}
\end{figure}

\begin{figure}[h!]
\centering
\includegraphics[width = 0.4\textwidth]{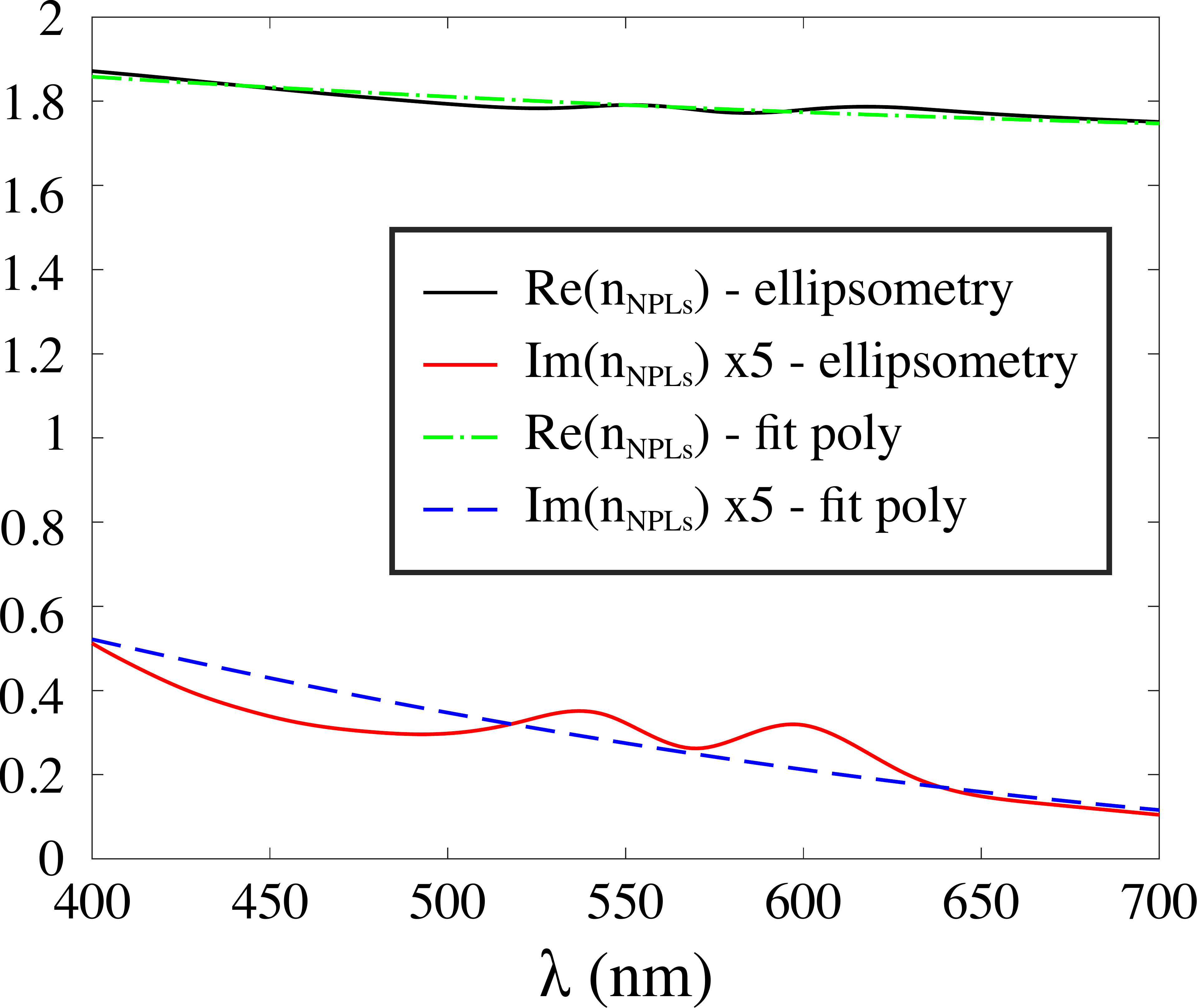}
\caption{Refractive index of the NPLs measured by ellipsometry and comparison with the polynomial fit whose coefficients are given in Table \ref{tab_coeff_indice}, from a sample composed of $42 \pm 5$ nm thick layer of NPLs deposited by spin coating on top of a 50 nm of silver on a 1 nm of Germanium, on a SF10 glass substrate. For the sake of clarity, the imaginary part of the refractive index is multiplied by 5.}\label{indice_Npls}
\end{figure}

In order to measure the refractive index of the NPLs, we fabricated a sample consisting on NPLs deposited by spin coating on top of a stack 50 nm thick layer of silver/ 1 nm thick layer of germanium/ SF10 glass substrate. Silver and germanium were deposited using electron beam evaporation. To deposit the NPLs, 200 \textmu L of a NPLs solution was spin coated on the metallic substrate at 500 rpm during 30 seconds with an acceleration ramp of 5 seconds. The ellipsometry measurements were performed in three steps. 
\begin{enumerate}
    \item First, the refractive index of a SF10 glass substrate was measured to serve as a reference.
    \item The refractive index of silver (including 1 nm of germanium) was measured from a reference sample which was fabricated under the same conditions as the sample covered with NPLs. It thus consists in a 50 nm thick layer of silver on a 1 nm thick layer of germanium, on a SF10 glass substrate. The experimental refractive index is similar to the silver index of reference \cite{Palik12}, as it can be seen in Figure \ref{indice_Ag}, so that the germanium layer has little impact on the refractive index. The thickness of the layer was obtained by scratching it with a needle and measuring the depth of the slit by AFM. We obtained $50 \pm 3$ nm, in agreement with the nominal value.
    \item Knowing the refractive index models of glass and silver, the refractive index of the NPLs was extracted from ellipsometry data and processed using a B-spline method (which is Kramers-Kronig consistent). The refractive index is given in Figure \ref{indice_Npls}. The total thickness of the sample was obtained by scratching it with a needle and measuring the total depth of the slit by AFM. By substracting the experimental thickness values of the silver and germanium layers obtained from the reference sample, we obtained $42 \pm 5$ nm thick. 
\end{enumerate}

\newpage
In order to perform the dispersion relation computed with a complex frequency presented in Figure 5 in the main article \cite{article}, we fitted the index of the NPLs as well as the index of silver by a polynomial of degree 2: $p(\lambda) = p_{1}\lambda^{n} + p_{2}\lambda^{n-1}+...+p_{n}\lambda+p_{n+1}$, with the Matlab$^{\mbox{\scriptsize{\textregistered}}}$ function "polyfit". The fitting coefficients are given in Table \ref{tab_coeff_indice}. The comparisons between the ellipsometry measurements and the polynomial fits are presented in Figure \ref{indice_Ag} for the silver and in Figure \ref{indice_Npls} for the NPLs. A higher degree of the polynomial fits more accurately the experimental data, but the dispersion relation remains the same. \\
Nevertheless, the absorptivity computations presented in the main article \cite{article} are done with an interpollation of experimental index of the NPLs obtained by ellipsometry and the refractive index of silver of reference \cite{Palik12}.

\begin{table}[h!]
    \centering
    \begin{tabular}{c|c|c}
    & NPLs & Ag\\
        $p_{1}$ &  $0.5186 + 0.3912i$ & $1.0159 - 3.7984i$\\
        $p_{2}$ &  $-0.9402 - 0.7008i$ & $-1.2376 +12.7632i$\\
        $p_{3}$ &  $2.1514 + 0.3220i$ & $0.5036 - 2.5394i$\\
    \end{tabular}
    \caption{Coefficients for the polynomial fit $p(\lambda) = p_{1}\lambda^{n} + p_{2}\lambda^{n-1}+...+p_{n}\lambda+p_{n+1}$, of degree $n=2$ of the refractive index of the nanoplatelets (NPLs) of Figure \ref{indice_Npls} and the refractive index of silver (Ag) of Figure \ref{indice_Ag}.}
    \label{tab_coeff_indice}
\end{table}

\newpage
\section{Spatial structure of the surface plasmon}
\begin{figure}[h!]
\centering
\includegraphics[width = 0.6\textwidth]{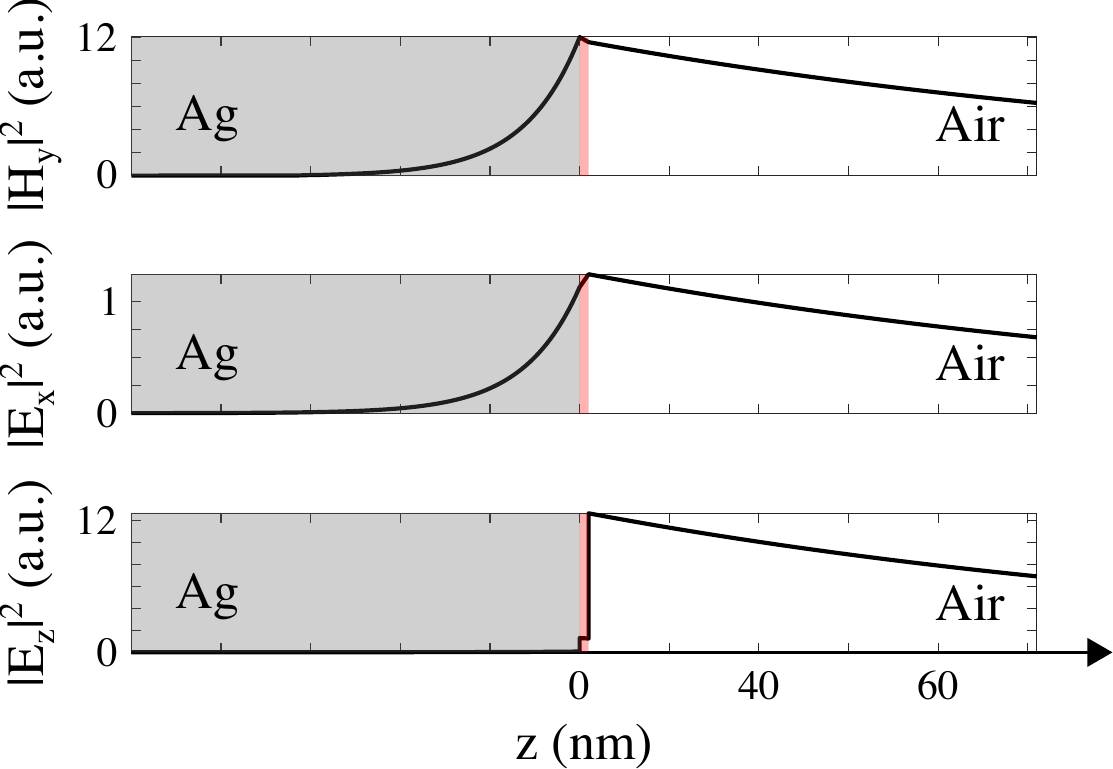}
\caption{Spatial structure of the mode as a function of z at 607.4 nm (in arbitrary units), for a sample composed of 2 nm thick layer of NPLs (in pink hue area) on silver. At this wavelength the refractive indexes are $n_{\mathrm{NPLs}}=1.7844 + 0.0598i$, $n_{\mathrm{Ag}}=0.1260 + 3.7852i$ and $n_{\mathrm{eff}}=1.0419 + 0.0031i$.}\label{fig_champ}
\end{figure}

This section shows the spatial structure of the mode which exists at the interface between a silver substrate and a thin layer (2 nm) of NPLs at 607.4 nm, computed with the refractive index of the NPLs measured by ellipsometry and the refractive index of silver of reference \cite{Palik12}. Figure \ref{fig_champ} shows that the mode is evanescent both in metal and in air. It corresponds to a surface plasmon polariton at the interface metal/NPLs/air.

\newpage

\section{Estimation of the beam efficiency}
\begin{figure}[h!]
\centering
\includegraphics[width = 0.7\textwidth]{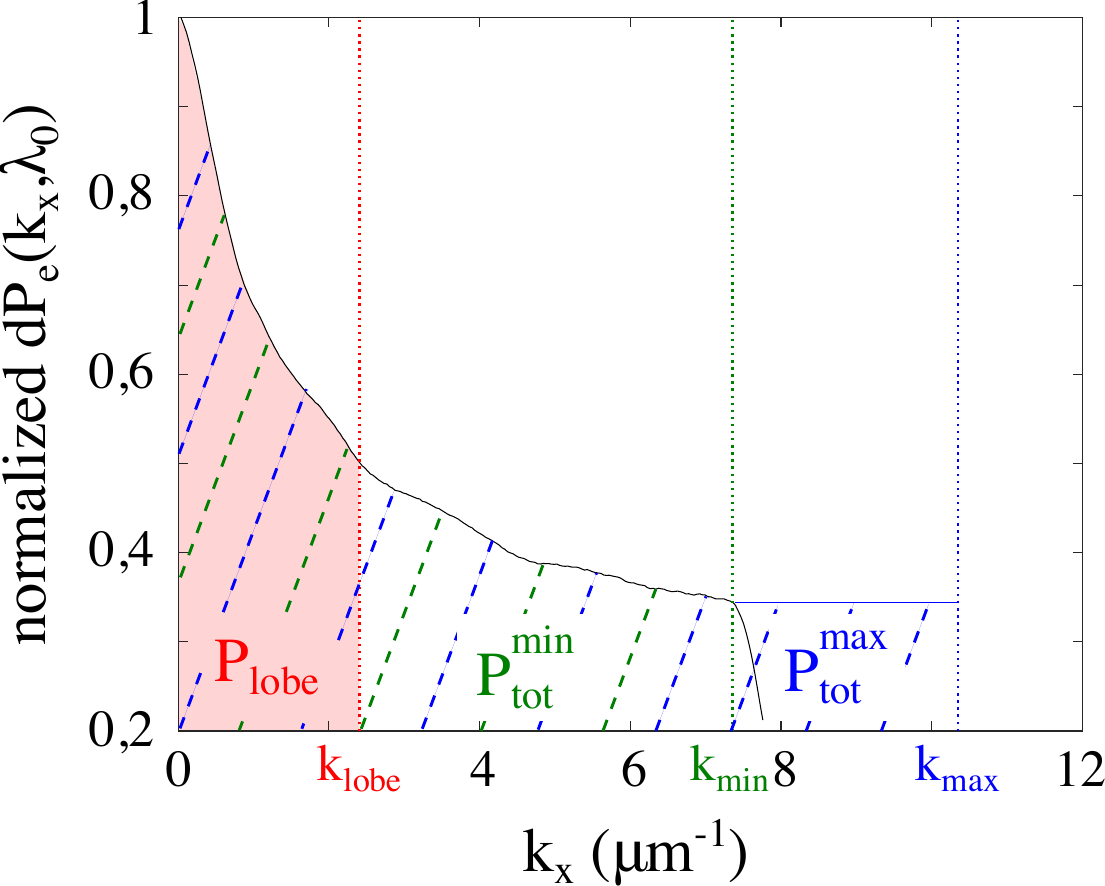}
\caption{Normalized experimental emitted power density as a function of $k_{x}$ at the peak emission wavelength, for the 2D silver metasurface covered with NPLs. Three emitted power values, each defined by different wavevector integration boundaries, can be computed: $P_{\mathrm{lobe}} = \int_{0}^{k_{\mathrm{lobe}}} \mathrm{d}P_{\mathrm{e}}$, $P^{\mathrm{min}}_{\mathrm{tot}} = \int_{0}^{k_{\mathrm{min}}} \mathrm{d}P_{\mathrm{e}}$, $P^{\mathrm{max}}_{\mathrm{tot}} = \int_{0}^{k_{\mathrm{max}}} \mathrm{d}P_{\mathrm{e}}$, with $k_{\mathrm{lobe}} = 2.40$ \textmu m$^{-1}$, $k_{\mathrm{min}} = 7.36$ \textmu m$^{-1}$ and  $k_{\mathrm{max}} = 10.35$ \textmu m$^{-1}$.}\label{fig_eff}
\end{figure}

In this section, we present the computation of the beam efficiency at the experimental peak emission wavelength (at 607.4 nm), defined as $P_{\mathrm{lobe}}/P_{\mathrm{tot}}$, where 
$P_{\mathrm{lobe}} = \int_{0}^{k_{\mathrm{lobe}}} \mathrm{d}P_{\mathrm{e}}$ is the emitted power in the emission peak represented in red in Figure \ref{fig_eff}, with $k_{\mathrm{lobe}} = 2.40$ \textmu m$^{-1}$.\\
Since the signal is symmetrical in $\pm k_{x}$, we integrate over the positive axis only and multiply by two.\\
The total power $P_{\mathrm{tot}}$ has been determined in two different ways:
\begin{itemize}
    \item Experimentally, light is collected within a light cone limited by the numerical aperture of the objective (NA = 0.75), so that it is not possible to obtain the exact total power emitted between 0$^{\circ}$ and 90$^{\circ}$. However, it is possible to estimate the beam efficiency with the total power collected, called $P^{\mathrm{min}}_{\mathrm{tot}} = \int_{0}^{k_{\mathrm{min}}} \mathrm{d}P_{\mathrm{e}}$, represented with green dotted lines in Figure \ref{fig_eff}. We chose a value of $k_{\mathrm{min}} = 7.36$ \textmu m$^{-1}$ (corresponding to 45.3$^{\circ}$ at 607.4 nm), slightly lower than $k_{\mathrm{NA}} = k_{0}\mathrm{NA}$, before the signal decreases (see Fig. \ref{fig_eff}). This estimation gives an overestimation of the beam efficiency value. We obtain a beam efficiency of 44.8 $\%$. \\
    \item It is also possible to give a boundary value of the total emitted power emitted between 0 and 90$^{\circ}$, by extrapolating the value of the emitted power at $k_{\mathrm{min}}$ for $k>k_{\mathrm{min}}$. The integrated power is then $P^{\mathrm{max}}_{\mathrm{tot}} = \int_{0}^{k_{\mathrm{max}}} \mathrm{d}P_{\mathrm{e}}$ and is represented with blue dotted lines in Figure \ref{fig_eff}. Thus, we obtain an underestimation of the beam efficiency value of 35 $\%$.
\end{itemize}
We therefore estimate that the radiative efficiency lies in the range 35\% and 44.8 \%.

\newpage
\section{Emission and Absorption for TE and TM polarization states}

\begin{figure}[h!]
\centering
\includegraphics[width = 1\textwidth]{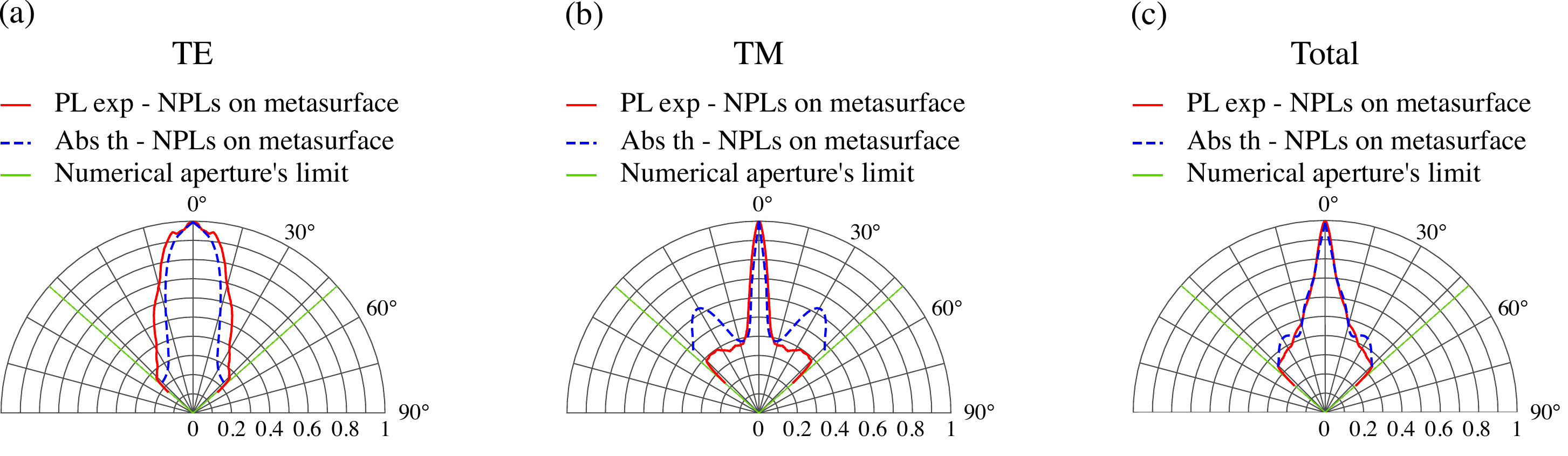}
\caption{Normalized experimental radiation patterns at their peak emission wavelengths (red curve) and normalized theoretical absorption pattern calculated at the same peak emission wavelengths (blue dotted line), as a function of the polarization state (a): Transverse Electric (TE) at 608.4 nm, (b): Transverse Magnetic (TM) at 605.8 nm, (c): Total emission at 607.4 nm, for  $h_{\mathrm{res}} = 100$ nm, $l_{\mathrm{res}} =$ 450 nm, $p_{\mathrm{res}} = $ 600 nm, $h_{\mathrm{top-NPLs}} =$ 2 nm.}
\label{htop2}
\end{figure}

We present in Figure \ref{htop2} the comparisons between normalized experimental radiation patterns and the normalized absorptivities, for Transverse Electric (TE), Transverse Magnetic (TM) polarization states, and for the total emitted power, plotted at their experimental peak emission wavelengths. Only the last case is presented in the article \cite{article}.

\begin{figure}[h!]
\centering
\includegraphics[width = 1\textwidth]{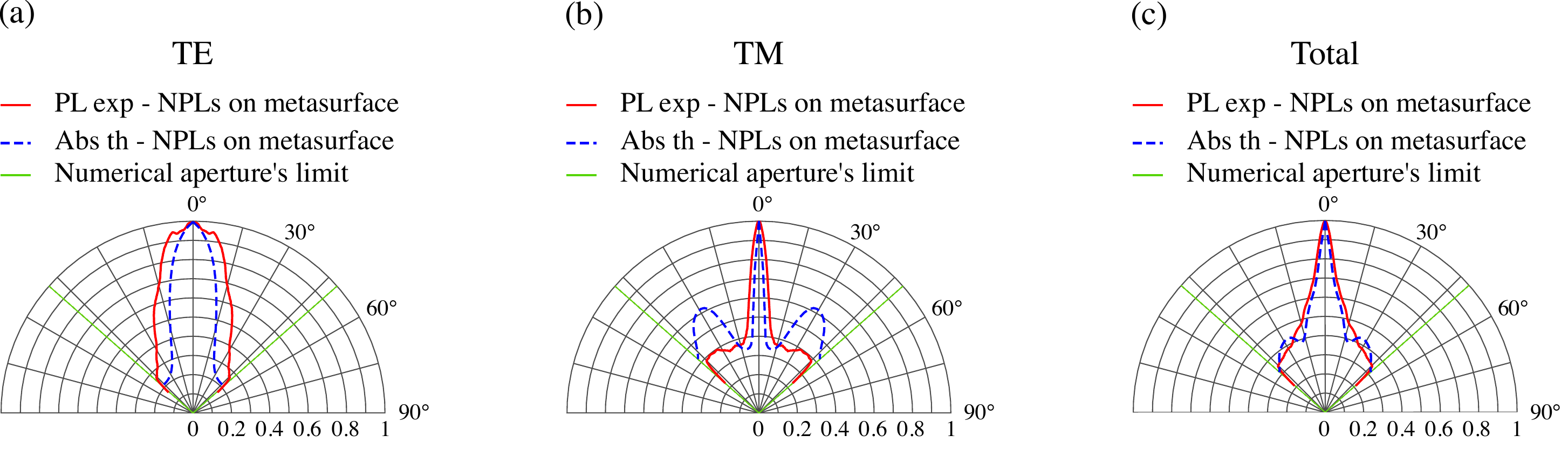}
\caption{Normalized experimental radiation patterns at their peak emission wavelengths (red curve) and normalized theoretical absorption pattern calculated at the same peak emission wavelengths (blue dotted line), as a function of the polarization state (a): Transverse Electric (TE) at 608.4 nm, (b): Transverse Magnetic (TM) at 605.8 nm, (c): Total emission at 607.4 nm, for  $h_{\mathrm{res}} = 100$ nm, $l_{\mathrm{res}} =$ 450 nm, $p_{\mathrm{res}} = $ 600 nm, $h_{\mathrm{top-NPLs}} =$ 0 nm.}
\label{htop0}
\end{figure}

We also present in Figure \ref{htop0} the comparison between the experimental radiation patterns and the normalized absorptivities for $h_{\mathrm{top-NPLs}} = 0$  nm, that is considering there was no overfilling of the grating grooves, and showing a less good agreement.

\newpage